\documentclass[openacc]{rstransa}

\begin{document}

\title{The Standard Model: How far can it go and how can we tell?}

\author{J. M. Butterworth$^{1}$}

\address{$^{1}$Department of Physics \& Astronomy, University College London}

\subject{Maxwell, Particle Physics}


\corres{Jon Butterworth\\
\email{J.Butterworth@ucl.ac.uk}}

\begin{abstract}
The Standard Model of particle physics encapsulates our current best understanding of physics at the smallest 
distances and highest energies.
It incorporates Quantum Electrodynamics (the quantised version of Maxwell's electromagnetism) and the weak and 
strong interactions, and has survived unmodified for decades, save for the inclusion of non-zero neutrino masses 
after the observation of neutrino oscillations in the late 1990s. It describes a vast array of data over a wide range of 
energy scales.
I review a selection of these successes, including the remarkably successful prediction of a new scalar boson, 
a qualitatively new kind of object observed in 2012 at the Large Hadron Collider. New calculational techniques 
and experimental advances challenge the Standard Model across an ever-wider range of phenomena, now extending 
significantly above the electroweak symmetry breaking scale. I will outline some of the consequences of these 
new challenges, and briefly discuss what is still to be found.
\end{abstract}


\begin{fmtext}
\section{Introduction}

Consider a naive question. If take any object and cut it into half, then cut it in half again, and again, 
and keep doing that, what to I get in the end? What structure is revealed?
Indeed, do I reach an end, or can I carry on for ever? 

Answering this question is one of the goals of physics. 
It is reductionist approach which is not the whole story, of course - whatever tiny constituents are revealed,
their interactions lead to rich emergent phenomena revealing new physics, not to mention chemistry, biology 
and the rest. However, knowledge of the structure of matter at the smallest accessible scales is important, and is 
surely one of the most exciting frontiers of science.

\end{fmtext}
\maketitle

The resolution required to see the ever-smaller pieces requires higher and higher energy, and thus the
frontier of the very small becomes the high-energy frontier. Similarly, the energies and distances involved
map the same frontier on to the physics of the very early universe; the hot, dense moments after the big bang.
But in the end, the question is the same - what is the universe made of?

The current answer is encapsulated in the Standard Model. 
Quarks in six flavours - up, down, charm, strange, top and bottom. 
Leptons - also six kinds, the electron, muon and tau, plus three neutrinos. 
There are also corresponding antiparticles. 
These interact with each other by exchange of gauge bosons - the photon carrying electromagnetism, the 
$W$ and $Z$ bosons carrying the weak force, and gluons carrying the strong force. 
And in the background lies the Higgs boson.

Like any most answers in science, this is a provisional statement, 
determined by our current ability to halve things, or, less destructively, 
to resolve ever-tinier objects. But it has just achieved a major predictive success with
the discovery of the Higgs boson, and it is remarkable that we have a self-consistent theory 
in which these objects are pointlike, 
and which can describe phenomena over an enormous range of energy and distance scales. In the following
sections I will recount some selected highlights of the model, and look at where further progress may be 
expected.

\section{Atomic scales}

I will start the tour of the Standard Model at typical atomic scales of around $10^2$ eV.
(The gluon and the photon masses, at zero, are off the low end of a logarithmic scale, and neutrino masses 
I will return to later.) Precision atomic 
physics measurements played a critical role in the development of the QED sector of the Standard Model, and the
current poster-child for high precision quantum-field theory is the anomalous magnetic momentum of the electron,
where theory and experiment are in agreement at the level of one part in $10^{13}$~\cite{Odom:2006zz,Agashe:2014kda}.
Obtaining this precision in the theory requires the incorporation of electroweak and strong corrections, 
due to contributions to the interaction between the electron and the photon coming from loops involving virtual
particles. Since non-Standard Model particles may also contribute in such loops, this measurement, coupled with 
precise measurements of the fine structure constant, can set a limits on some types of new physics models at 
energy scales much 
higher than atomic physics scales\cite{Giudice:2012ms}. The same is true of the
anomalous magnetic moment of the muon, and since many new physics contributions scale with the square of the 
lepton mass,
the sensitivity of these measurements to such contributions is higher. The muon anomalous magnetic moment is less 
precisely known than that of the electron (though still to an 
impressive accuracy of about one part in $10^{10}$), and currently exhibits a 3.6$\sigma$ discrepancy between theory
and measurement. Taken at face value, this is evidence for physics beyond the Standard Model, potentially 
at scales close to the electroweak scale ($10^{11}$ eV). New measurements and improved calculations will 
hopefully soon clarify 
the situtation\cite{Grange:2015fou}.

\section{QCD and hadron masses}

Moving upward in energy scale from atomic binding energies, past the electron mass ($0.5 \times 10^{6}$ eV) and 
nuclear binding energies ($\approx 10^7$ eV), we reach an energy scale of great importance, 
$\Lambda_{QCD} \approx 2 \times 10^{8}$ eV. For a specified renormalisation scheme, this scale is equivalent 
to the dimensionless strong coupling at some reference scale, and is a fundamental parameter of QCD. It determines the
scale at which quarks and gluons are confined inside hadrons. In a sense it is only above this scale that quarks
and gluons become the relevant degrees of freedom in the Standard Model. Below this scale, they cannot be resolved 
inside hadrons, which are (at least mostly) $q\bar{q}$ or $qqq$ bound states.

In this region the strong coupling is still too large to allow the convergence of a perturbative expansion in terms of
Feynman diagrams. However, lattice techniques can be used to calculate important emergent physical properties such as
the confinement potential and hadron masses (see for example~\cite{Bali:2000gf,Aoki:2013ldr}). Recent experimental 
progress in exploring the hadronic mass spectrum and confinement includes the discovery of 
tetraquark ($qq\bar{q}\bar{q}$) and 
pentaquark ($qqqq\bar{q}$) states\cite{Choi:2007wga,Chilikin:2013tch,Aaij:2014jqa,Aaij:2015tga}.

\section{`Hard' QCD}

As the energy scale increases well above $\Lambda_{QCD}$ and typical hadron masses, 
the coupling constant, $\alpha_s$, decreases, and the strong force exhibits the property of 
asymptotic freedom\cite{Gross:1973id,Politzer:1973fx}. 
This means that at scales of around $10^{10}$ eV and above, quarks and gluons are indeed the relevant 
degrees of freedom, and calculations can be performed using perturbative expansions in $\alpha_s$.

The two most important and direct ways in which the existence and behaviour of quarks and gluons
can be explored in this energy regime are in the production of hadronic 'jets', and the study of 
the internal structure hadrons. And as well shall see, both types of measurement are interconnected.

Jets result from a short-distance, high-energy scatters between a pair of partons (a general terms used 
to refer to quarks and gluons), or from the decay of a massive particle into partons. Jets are seen
for example in the process $e^+e^- \rightarrow$ hadrons, for centre-of-mass energies of more than a few 
$10^9$ eV, where the underlying process at leading order in the perturbative expansion is taken to
be the electroweak annihilation of the electron and positron to a virtual photon ($\gamma$) or $Z$ boson, 
which then decays to $q\bar{q}$. While weakly coupled at short distances, those partons will, as they 
fly apart, experience the linearly increasing confinement potential mentioned in the previous section.
The increasing potential energy with distance implies that it is energetically favourable to 
create more partons, shortening the distances over which the potential acts; a process ending at lower 
energies in the production of color-neutral hadrons. These hadrons form collimated jets which 
preserve, to a great extent, the kinematics of the initiating partons. The production of three-jet 
events\cite{Ellis:1976uc}
at the Petra collider in DESY\cite{Brandelik:1979bd} was generally seen as the first 
direct evidence for the existence of 
gluons, as the kinematics and cross sections were as predicted by calculations of the process
$e^+e^- \rightarrow q\bar{q}g$. Jets have been measured precisely in $e^+e^-, ep, p\bar{p}$ and $pp$ 
collisions and provide a stringent test of QCD and a sensitive probe of short-distance physics.

The first evidence for pointlike objects inside hadrons - later identified with the quarks already 
introduced by Gell-Mann and Zweig to explain hadron properties - came from the deeply inelastic scattering
of electrons off protons. The squared four-momentum transfer, $Q^2$, in such collisions dictates the resolution,
and it was seen that at some point, corresponding to energies well above $\Lambda_{QCD}$, the cross
section approximately 'scaled'. That is, once the propagator dependence was accounted for, the dependence 
on $Q^2$ was very weak, as expected if a pointlike particle has been resolved. The residual dependence on
$Q^2$ - scaling violation - is due to the radiation of gluons from the quarks, and, since it takes 
place at short distances, is calculable in QCD using perturbative techniques~\cite{Gribov:1972ri,Gribov:1972rt,Lipatov:1974qm,Dokshitzer:1977sg,Altarelli:1977zs}: the 
excellent agreement between data - principally from the HERA $ep$ collider\cite{Abramowicz:2015mha} 
- and fits incorporating 
these scaling violations to next-to-next-leading 
order in perturbative QCD is another vindication of the theory. 

The parton densities extracted in this way are needed to make predictions for jet (and other) cross 
sections in hadron colliders. And precise measurements of jets are used to constrain those parton 
densities further.

\section{The Electroweak scale}

The transition to the next important energy scale in physics is nicely illustrated by staying with deep
inelastic scattering. Electron-proton collisions can not only be mediated by photon exchange, but also
by the exchange of weak bosons, the $W$ and the $Z$. Since the $W$ carries electric charge, those
events will have distinct final states, in which the emerging scattered lepton has turned into a 
neutrino. As might be expected, the rates of these 'charged-current' events are much lower 
than those of the electromagnetic scatters, reflecting the relative strength of the electromagentic and
weak forces. However, as the energy scale increases, the rates converge, until at around $10^{11}$ eV,
they become roughly equal. This effect, measured at HERA\cite{Abramowicz:2015mha}, reflects the fact that most of the 
apparent difference in strength of 
the two forces at lower energies is due to the different propagator masses (zero for the photon, 
$~10^{11}$ eV for the $W$ and $Z$). Once the energy is such that these masses are no longer important,
a symmetry is restored, and the forces become comparable in strength. This energy scale is
known as the electroweak symmetry breaking scale.

\subsection{The $Z$ and the neutrinos}

Studies of the $Z$ boson, resonantly produced in $e^+e^-$ collisions at LEP and SLC\cite{ALEPH:2005ab}, 
provide the precise information on many features of electroweak physics. One major triumph 
was the precise measurement of the decay width of the $Z$; that is, the shape of the resonance. 
This can be precisely calculated in electroweak theory, and is sensitive to all the possible 
decay channels of the $Z$ including, crucially, the invisible decay channels $Z \rightarrow$ neutrinos.
These measurements indicate that there are three, and only three, flavours of neutrinos - and thus only
three generations of matter, assuming each generation contains a light, weakly-interacting neutrino.

\subsection{Loops and fits}

To realise the full potential of the precise data from LEP and LEP2, higher order corrections 
involving virtual-particle loops have to be implemented in the electroweak theory. This means that
the data have indirect sensitivity to the existence and properties of undiscovered particles, 
too heavy to be produced directly in the collisions. At the time of the experiments, such undiscovered
particles included the top quark. The discovery of the top quark at the Tevatron\cite{Abe:1995hr,Abachi:1995iq}, 
in a region consistent
with the Standard Model fits, switched attention to the last remaining unknown particle in the Standard Model,
the Higgs boson.

The Higgs boson is scalar particle predicted by the mechanism for electroweak symmetry breaking 
based on work in the early 1960s\cite{Englert:1964et,Higgs:1964ia,Higgs:1964pj,Guralnik:1964eu,Higgs:1966ev,Kibble:1967sv} 
which was later incorporated into the SM\cite{Glashow:1961tr,Weinberg:1967tq,salam,'tHooft:1972fi}. This mechanism allows 
the $W,Z$ and the fermions to acquire mass by coupling to a scalar field with a non-zero vacuum 
expectation value; introducing mass in this way preserves the gauge symmetries of the theory, and
thus the property of renormalizability required to make the theory predictive at higher orders and
higher energies.

The precise measurements of the top and $W$ masses at the Tevatron\cite{Aaltonen:2012ra,Aaltonen:2013iut} 
increased the power of the SM fits,
and - again via loop corrections - led to constraints on the mass of the SM Higgs boson,
should it in fact exist. Meanwhile, direct searches at LEP2, the Tevatron and early LHC data excluded 
signficant ranges of possible mass values.

\subsection{Discovery and vindication}

The discovery of the Higgs boson\cite{Aad:2012tfa,Chatrchyan:2012xdj}, 
with a mass ($~1.25 \times 10^{11}$) eV\cite{Aad:2015zhl} once more consistent with 
the constraints from electroweak precision data, was a remarkable triumph not only for the 
accelerator and the experiments, but for the 
theoretical ideas behind the SM. So far, all the properties of the boson - charge, spin, parity, and the
production cross section $\times$ branching ratio for those decay modes which have been measured - are
in agreement with expectations of the SM. The SM now has no ``missing particles''.
Without the experimental discovery of
this qualitatively unique object, the SM could have been just an effective theory valid 
only up to $10^{12}$ eV or so. With the knowledge that the Higgs boson exists, the theory 
is capable of making predictions at energies far above the electroweak scale. 

\section{To the TeV scale and beyond}

Armed with this new discovery and increased energy at the LHC, particle physics is now exploring 
multi-TeV scale physics - $10^{12}$ eV and beyond. Making predictions with the SM in this qualitatively
new regime brings new challenges for the theory.

\subsection{New features, new demands}

As the energy of a collision increases, so does the phase space accessible for the production of 
high-energy objects such as high-momentum jets, leptons and photons. Increasing the energy far
above the electroweak scale means that objects with masses around the electroweak scale can also 
be produced with high energies and/or in high numbers. In the SM this means top quarks and the $W, Z$ 
and Higgs bosons. Describing with any degree of precision the final states thus produced requires 
innovative calculational techniques. In particular, there is a need for the theory to be able to
reproduce what is actually measured by the experiment. This means that calculations of 'inclusive' quantities,
which integrate over large regions of phase space, are not sufficient, since the experiments must place 
selection requirements on the final state. For example, the detectors have a limited angular (rapidity) 
acceptance, and also cannot measure jets or particles at arbitrarily low transverse momenta. If the theory
can only calculate total cross sections, or totally inclusive quantities, then exptrapolations have to be
applied to the data in order to make comparisons. These extrapolations must make some theoretical assumptions
- typically they will use a Monte Carlo event generators, which simulate the fully exclusive final 
state\cite{Buckley:2011ms}. The
accuracy of the comparison is likely then not limited by the inclusive theory calculation, or by the intrinsic 
experimental uncertainty, but by the accuracy of the generator, as well as by the assumption that no unexpected 
physics occurs in the regions which are not covered by the detector.

For these reasons, much activity in the theory community is now focussed on precise exclusive calculations, 
which in many cases are implemented within Monte Carlo generators. This is not done so much to improve the
accuracy of the extrapolations, but to remove the need for them altogether, by making precise predictions
of realistic measurements within the fiducial acceptance regions of the detectors.

A key issue here is the matching of two perturbative approaches to QCD. The matrix element calculations 
at the heart of event generators can be increased in multiplicity, thus filling phase space more accurately,
and can also incorporate the full complement of higher-order terms, which means also including diagrams
involving loops (similar to those implicated in the muon magnetic moment, discussed above). 
In the latter case the precision
of the calculation is systematically improved.
There are, however, regions of phase space which contain kinematic enhancements which mean such an approach
converges very slowly, if at all. Such enhancements typically occur when partons are soft or almost colinear - 
for example in the early stages of jet formation.
They usually occur in such a way that the emission of a new parton introduces a large logarithm
of the ratio of some kinematic factors, which can negate the suppression arising from the introduction
of an additional $\alpha_s$ factor, and imply that the series will not converge. Fortunuately, techniques have
been developed to cast such terms in the form of summable series; they have also be implemented algorithmically
in Monte Carlo generators in the form of 'parton showers', which are essential to obtain accurate descriptions
of the final state.

The development of methods of matching these parton shower terms to higher multiplicity\cite{Catani:2001cc} 
and higher order\cite{Frixione:2002ik,Hamilton:2013fea}
matrix elements are breakthroughs without which much of the LHC programme would be seriously hampered. Advances
continue, and the importance of these developments is likely only to increase as colliders extend the energy
reach over which the SM can be confronted with data.

\subsection{Precision Higgs Physics}

Given the theoretical advances outlined above, the experiments can increasingly concentrate on making
precise and detailed exclusive and differential measurements. For example, ATLAS and CMS have already 
moved into the era of precision Higgs physics, not only measuring inclusive properties, but also differential 
fiducial cross sections. The multiplicity and distribution of jets accompanying the Higgs are
used to obtain separate sensitivity to the various production mechanisms (e.g. between gluon fusion and vector
boson fusion) and is intrinsic to the selection of some types of event. Measuring the transverse momentum
distribution of the Higgs boson production offers a more rigourous probe of the SM than measuring the integrated 
total. Several of these measurements have already been made (see for example \cite{Aad:2014lwa,Khachatryan:2015yvw}), 
and the number, reach and precision will
increase as long as the LHC continues to provide data.

\subsection{Boost and jet substructure}

Since soft QCD takes over somewhere around the $10^9$~eV region, and the energy scales involved in jet 
production may extend up to several $10^{12}$~eV, much of the evolution of the substructure of the jet 
takes places within the perturbative regime, and again can be predicted using the approaches described above. 
Jet substructure has increased relevance when the jet scale lies above the electroweak scale, 
since particles with masses around the 
electroweak scale ($W,Z$ and Higgs bosons and the top quark) may now be produced with high Lorentz boosts, 
and thus their decay products will be highly collimated and, in the case of hadronic decays, may be reconstructed 
as a single jet\cite{Seymour:1993mx,Butterworth:2002tt,Butterworth:2008iy,Kaplan:2008ie}. 
Such configurations have the advantage that combinatorial and other backgrounds can be suppressed, 
but they require the study of the substructure of jets in order to identify and reconstruct the 
decaying massive particle (see \cite{Altheimer:2013yza} for a review). 
Techniques based on these ideas have been widely employed in measurements at 
the LHC (for example of the top quark transverse momentum\cite{Aad:2015hna}), and in searches both for the SM 
Higgs decaying to b-quarks and probes of the SM (and potentially beyond) close to the kinematic limit.

\section{Flavour physics}

Omitted from the discussion so far, because it does not fit neatly into the progress of increasing energy
scale I have been following, is flavour physics. The masses of the
different flavours of fundamental fermions in the SM span a huge range, from the neutrinos at around $10^{-1}$ eV
to the top at $1.7 \times 10^{11}$ eV. The reasons for this are not contained within the SM itself. In addition, the
mass eigenstates of the particles are not identical to their (flavour) eigenstates under the weak interaction, 
meaning that
a unitary matrix with four free parameters is introduced in both the quark and lepton sectors. This leads to
the phenomenon of 'mixing' - the flavour of quarks and leptons, which is defined as the weak eigenstate, can 
change as the propagate, mediated by the mass eigenstates. This matrix also encodes the possibility of the 
simultaneous violation of the Charge and Parity symmetries, CP violation.

\subsection{Quarks}

Quark mixing and CP violation in the SM are described by the Cabbibo-Kobayashi-Masakwa 
(CKM) matrix. Measurements in $e^+e^-$ collisions at the at the 
B-factories, Babar and Belle\cite{Bevan:2014iga}, showed that the 
CP violation observed in the quark sector is, at least to a very 
good approximation, consistent with the mechanism allowed in the SM.
Investigations of quark flavour 
mixing at LHCb continue to probe the SM and increase the precision on the elements of the CKM matrix.

\subsection{Neutrinos}

Neutrinos were massless in the original SM, and the observation of neutrino mixing, which implies that they have 
non-zero mass, is the only fundamental change that has been imposed by data on the SM since its inception.
Neutrino oscillations were first pointed to by the observed deficit of electron-neutrinos coming from the 
Sun\cite{Cleveland:1998nv,Bahcall:2000nu}, 
and were directly observed in atmospheric and  Solar neutrino measurement by Super-Kamiokande\cite{Fukuda:1998mi} 
and SNO\cite{Ahmad:2002jz} respectively. 
The matrix which is introduced to describe the mixing is known as the Pontecorvo-Maki-Nakagawa-Sakata (PMNS) matrix,
and its elements are now being measured with increasing precision by accelerator and reactor-based neutrino 
experiments\cite{Abe:2008aa,Adamson:2011ig,An:2012eh,Abe:2014ugx}. 
It is known that, unlike the CKM matrix, the off-diagonal elements are relatively large. That is, unlike the quark case 
where the flavour and mass eigenstates are almost aligned, the neutrino flavour
eigenstates are an almost maximal mixture of the mass eigenstates. It is not yet known whether CP violation
occurs in the neutrino sector - the relevant parameter of the PMNS matrix has not yet been measured.

\section{Conclusion}

With a small number of fundamental objects and principles, the Standard Model describes and predicts an
enormous variety of physical phenomena, over energy scales ranging from zero up to several $10^{12}$ eV and 
potentially far beyond. Some of the principles behind the theory trace their origins back to Maxwell's
equations for electromagnetism, and those equations remain the classical form of the quantised electromagnetic 
force within the theory - QED. QED itself, as a gauge theory, is the template for the weak and strong forces,
themselves based on larger, non-Abelian gauge symmetries. The prediction that a scalar boson would appear at
some scale below $10^{12}$ eV, as a consequence of reconciling gauge symmetries and massive bosons, has been 
spectacularly vindicated, and extends the region of applicability of the Standard Model potentially up to
energies beyond the reach of particle colliders.

Before getting too pleased with ourselves at the power and subtlety of the SM, (or despairing of our 
ability to extend it) it is worth remembering that is fails entirely to incorporpate one of the fundamental 
forces. Gravity is described by the general theory of relativity, and in that context is a consequence 
of the curvature of space-time rather than a force like those in the SM, mediated by gauge bosons. 
This is a very significant failure in the ambition to describe natural phenomena in a single framework. Worse, even in partnership
with general relativity, the rotation curves of galaxies, weak lensing of light, and the large scale structure of
the universe cannot be described without introducing 'Dark Matter', for which the SM does not contain a compelling
candidate (though the possibility of strongly-bound composite SM states is still discussed, 
see for example \cite{Jacobs:2014yca}). 
The low level of CP violation
possible from SM mechanisms has not been reconciled with the gross violation inherent in the fact that
the observable universe seems to consist overwhelmingly of matter rather than antimatter. And finally,
it is probably apparent from the preceding sections that the SM contains quite a lot of parameters with
seemingly arbitrary, but suggestive, values. These include some, such as the Higgs mass, that look
unnaturally 'fine tuned', to the extent that it is very tempting to think that they are fixed by some
so-far unknown symmetry or principle inherent in a larger theory, of which the SM is just a part.

So the story, of which Maxwell's work was such a important component, is not over. 
There is more to be found by continuing to look more and more closely 
at nature. We don't know whether any of the hints or inconclusive anomalies currently 
exercising particle physicists' minds 
(g-2, various bumps and  rare decays at the LHC, for example) will fade, or grow and lead to answers 
to some of the open questions above. But we do know that there are unknown territories
to be explored, and important answers to be found.

\vskip6pt

\ack{I'd like to thank Anatoly Zayats, John Ellis, Roy Pike, and the staff of the Royal Society 
for organising an excellent
meeting, and especially for the chance to see Maxwell's original manuscript\cite{Maxwell:1865zz}.}

\funding{Funded by UCL and STFC.}

\conflict{I have no competing interests.}


\end{document}